\documentclass[%
 aip,
 amsmath,amssymb,
 reprint,%
]{revtex4-1}

\usepackage{graphicx}% Include figure files
\usepackage{dcolumn}% Align table columns on decimal point
\usepackage{bm}% bold math
\usepackage[utf8]{inputenc}
\usepackage[T1]{fontenc}
\usepackage{mathptmx}
\usepackage{etoolbox}
\usepackage{color,soul}

\makeatletter
\def\@email#1#2{%
 \endgroup
 \patchcmd{\titleblock@produce}
  {\frontmatter@RRAPformat}
  {\frontmatter@RRAPformat{\produce@RRAP{*#1\href{mailto:#2}{#2}}}\frontmatter@RRAPformat}
  {}{}
}%
\makeatother
\begin{document}

\preprint{AIP/123-QED}

\title{Photoluminescence and charge transfer in the prototypical 2D/3D semiconductor heterostructure MoS$_2$/GaAs}
% Force line breaks with \\

\author{Rafael R. Rojas-Lopez}
   \email{rrlopez@fisica.ufmg.br}
\affiliation{Departamento de Física, Universidade Federal de Minas Gerais, 30123-970, Belo Horizonte, Brazil.}
\affiliation{Zernike Institute for Advanced Materials, University of Groningen, 9747 AG Groningen, The Netherlands.}
\author{Juliana C. Brant}
\affiliation{Departamento de Física, Universidade Federal de Minas Gerais, 30123-970, Belo Horizonte, Brazil.}
\author{Maíra S.O. Ramos}
\affiliation{Departamento de Física, Universidade Federal de Minas Gerais, 30123-970, Belo Horizonte, Brazil.}
\author{Túlio H.L.G. Castro}
\affiliation{Departamento de Física, Universidade Federal de Minas Gerais, 30123-970, Belo Horizonte, Brazil.}
\author{Marcos H.D. Guimarães}%
\affiliation{Zernike Institute for Advanced Materials, University of Groningen, 9747 AG Groningen, The Netherlands.}
\author{Bernardo R.A. Neves}
\affiliation{Departamento de Física, Universidade Federal de Minas Gerais, 30123-970, Belo Horizonte, Brazil.}
\author{Paulo S.S. Guimarães}
\affiliation{Departamento de Física, Universidade Federal de Minas Gerais, 30123-970, Belo Horizonte, Brazil.}

\begin{abstract}
The new generation of two-dimensional (2D) materials has shown a broad range of applications for optical and electronic devices. Understanding the properties of these materials when integrated with the more traditional three-dimensional (3D) semiconductors is an important challenge for the implementation of ultra-thin electronic devices. Recent observations have shown that by combining MoS$_2$ with GaAs it is possible to develop high quality photodetectors and solar cells. Here, we present a study of the effects of intrinsic GaAs, p-doped GaAs, and n-doped GaAs substrates on the photoluminescence of monolayer MoS$_2$. We observe a decrease of an order of magnitude in the emission intensity of MoS$_2$ in all MoS$_2$/GaAs heterojunctions, when compared to a control sample consisting of a MoS$_2$ monolayer isolated from GaAs by a few layers of hexagonal boron nitride. We also see a dependence of the trion to A-exciton emission ratio in the photoluminescence spectra on the type of substrate, a dependence that we relate to the static charge exchange between MoS$_2$ and the substrates when the junction is formed. Scanning Kelvin probe microscopy measurements of the heterojunctions suggest type-I band alignments, so that excitons generated on the MoS$_2$ monolayer will be transferred to the GaAs substrate. Our results shed light on the charge exchange leading to band offsets in 2D/3D heterojunctions which play a central role in the understanding and further improvement of electronic devices.
\end{abstract}

\maketitle

%\section*{Introduction}

Layered transition metal dichalcogenides (TMDs) are among the most studied two-dimensional (2D) materials in the last decade. Their atomically-thin structure and physical properties have attracted attention not only because of their interesting fundamental physics but also due to their potential applications for ultra-thin technological devices \cite{Yu2013, Nalwa2020, Sarkar2020, Pradeepa2020, Jia2019, Xu2016, Lin2015, Wang2012}. Similar to graphene, these materials can be mechanically exfoliated to obtain single layers. Of special interest are MoS$_2$, MoSe$_2$, WS$_2$ and WSe$_2$, TMDs that have been widely studied mostly because they suffer a transition from an indirect to a direct bandgap semiconductor when the monolayer thickness is achieved \cite{Mak2010,Splendiani2010}. As a consequence, the photoluminescence (PL) of the monolayer of these materials is much more intense when compared to that of the bulk material \cite{Splendiani2010}. Also, owing to their two-dimensional nature, TMD monolayers have their PL spectra dominated by excitonic effects. For monolayer MoS$_2$, a characteristic PL spectrum can usually be decomposed into three main peaks related to the recombination of different excitons, the so-called A and B excitons, and charged excitons, the trions \cite{Jadczak2017}. The large spin-orbit splitting (SOS) at the top of the valence band is responsible for the existence of the two exciton states, A and B \cite{Zhu2011,Splendiani2010}, while the third PL peak routinely observed in the PL spectrum of monolayer MoS$_2$ corresponds to charged A-excitons, or trions, which are tightly bound and are observed even at room temperature \cite{Mak2013a}.

In the monolayer limit, the properties of all TMDs are highly affected by the substrate on which they are deposited \cite{Sun2017, Buscema2014}. One interesting substrate for these monolayer materials is GaAs, a prototypical semiconductor which has been extensively studied and employed for electronics and optoelectronics applications that take advantage of its direct gap (1.42 eV at room temperature)  and relatively high electron mobility (up to 8000 cm$^2$ V$^{-1}$ s$^{-1}$ at room temperature) \cite{Sze2006}. The combination of the optical and electronic properties of TMDs and GaAs as a substrate has already shown promising results for implementation of solar cells \cite{Lin2015}, with a power conversion efficiency of up to 9.03$\%$, and photodetectors \cite{Xu2016,Sarkar2020,Jia2019,Zhang2017}, with a detectivity of up to 1.9 $\times$ 10$^{14}$ Jones. The success of these proof-of-concept studies urges the need to investigate in detail the properties of MoS$_2$/GaAs heterojunctions, in order to further improve device quality \cite{Padma2019}. Particularly, the band alignment between the two materials is still not well established although it is of major importance for applications involving these 2D/3D semiconductor architectures.

Here, we present a study of the effect of GaAs substrates on monolayer MoS$_2$ by analyzing the changes in the photoluminescence spectra of monolayer MoS$_2$ on GaAs substrates with different doping levels. We used three types of commercially-available GaAs substrates that we identify hereon as i-GaAs for intrinsic GaAs (semi-insulating), p-GaAs for Zn-doped p-type GaAs and n-GaAs for Si-doped n-type GaAs. The doping concentrations are $\sim$10$^{18}$ cm$^{-3}$ for both n-GaAs and p-GaAs. As a reference, we have control samples on two substrates, SiO$_2$/Si and n-GaAs, with the transferred MoS$_2$ monolayer isolated from the substrates by a bulk hexagonal boron nitride (hBN) flake. We propose a type-I band alignment, with a charge transfer between GaAs and the MoS$_2$ monolayer which depends on the GaAs doping. This band alignment model is supported by Scanning Kelvin Probe Microscopy (SKPM) measurements in the heterostructures.

%\section*{Experimental Methods}

Monolayers of MoS$_2$ (ML-MoS$_2$) were mechanically exfoliated and transferred to the substrates through the all-dry viscoelastic stamp method \cite{Castellanos-Gomez2014}. Similar processes were used to exfoliate and transfer the hBN bulk to the Si/SiO$_2$ and n-GaAs substrates. To confirm the single layer character of the MoS$_2$ flakes we used Raman spectroscopy to monitor the separation in frequency of the well-known $A_{1g}$ and $E^{1}_{2g}$ Raman modes \cite{Lee-Heinz-Hone-ACSNano2010,Li-Zhang-Advanced-Funct-Materials2012}, see Supplementary Information (SI).

The samples were studied in two sets. The first set was composed of a control sample of ML-MoS$_2$ on hBN/SiO$_2$/Si substrate (MoS$_2$/hBN/SiO$_2$) and three samples of ML-MoS$_2$ on GaAs with different doping:  MoS$_2$/i-GaAs, MoS$_2$/p-GaAs and MoS$_2$/n-GaAs. The second set is composed of two samples, one ML-MoS$_2$ on n-GaAs and one ML-MoS$_2$ control sample on hBN/n-GaAs substrate (MoS$_2$/hBN/n-GaAs). The second set of samples allowed us to verify the reproducibility of the results obtained for ML-MoS$_2$ as well as to produce a control sample with a dielectric environment that allows better comparisons of SKPM measurements made on different samples (see SI).

\begin{figure}[t!]
    \centering
    \includegraphics[width=\linewidth]{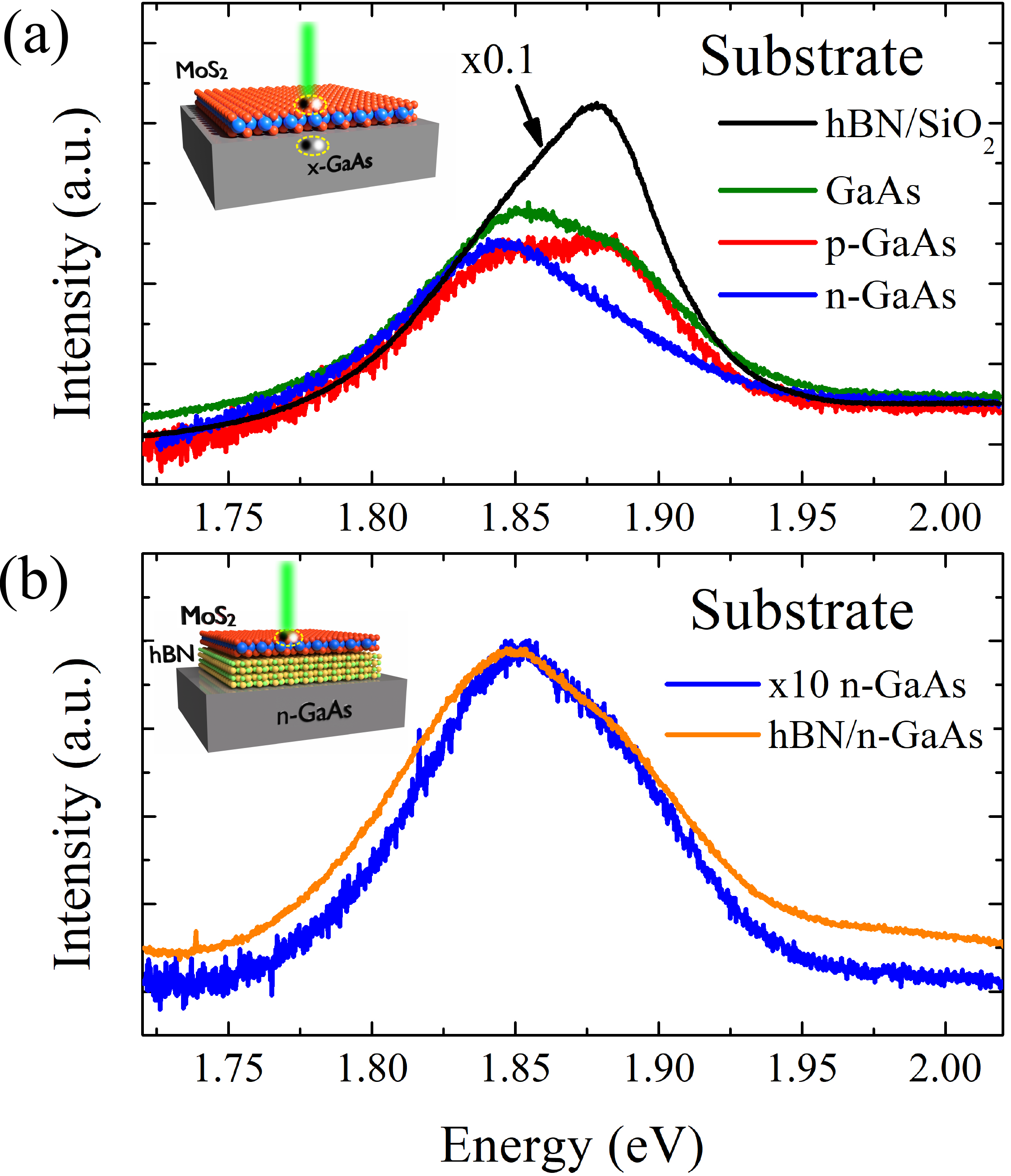}
    \caption{Photoluminescence spectra of ML-MoS$_2$ from the first set of samples (a) and from the second set of samples (b). Insets: Representation of the studied samples on x-GaAs substrate (a) and on hBN/n-GaAs substrate (b). }
    \label{Fig1:Samples and PL}
\end{figure}

We start our considerations by the PL measurements, which were accomplished with the same experimental conditions for all the samples. We are cautious with the laser exposure and spectra acquisition to minimize changes in the PL caused by photodoping effects\cite{Cao2021} and to allow the comparison of PL spectra from different samples (details are provided in the SI). The ML-MoS$_2$ spectra were obtained after removing the background photoluminescence from the GaAs substrate when applicable (see SI). In Figure \ref{Fig1:Samples and PL}a we present the ML-MoS$_2$ emission for the first set of samples. The intensity of the emission from ML-MoS$_2$ is approximately the same (within experimental resolution) for all MoS$_2$/x-GaAs (x=p, n, i) samples. However, their PL signals are around 10 times less intense than that of the ML-MoS$_2$ from the MoS$_2$/hBN/SiO$_2$ control sample. This observation suggests an important quenching mechanism for the ML-MoS$_2$ photoluminescence in the MoS$_2$/x-GaAs 2D/3D heterostructures, which is independent of the substrate doping level. We suggest two main paths for the reduction of PL from MoS$_2$ on GaAs: exciton dissociation through the junction and exciton transfer from MoS$_2$ to GaAs. The first process will contribute more if ML-MoS$_2$/x-GaAs form a type II heterojunction and the latter will be more important in a type I heterojunction. Therefore, we will try to elucidate the band alignment of the heterojunctions with other observations and the discussion that follows.

The results shown in Figure \ref{Fig1:Samples and PL}a are consistent with measurements on a second set of samples: MoS$_2$/hBN/n-GaAs and MoS$_2$/n-GaAs. We observe a 10:1 relation between the PL of the sample containing the hBN spacer to the one without this spacer (Figure \ref{Fig1:Samples and PL}b). This confirms that the hBN bulk layer worked well to isolate the ML-MoS$_2$ from the n-GaAs substrate, preventing exciton dissociation/transfer. From now, we are going to consider  just the control sample of the second set, as it presents a comparable dielectric environment with the first set of samples.

\begin{figure}
    \centering
    \includegraphics[scale=0.12]{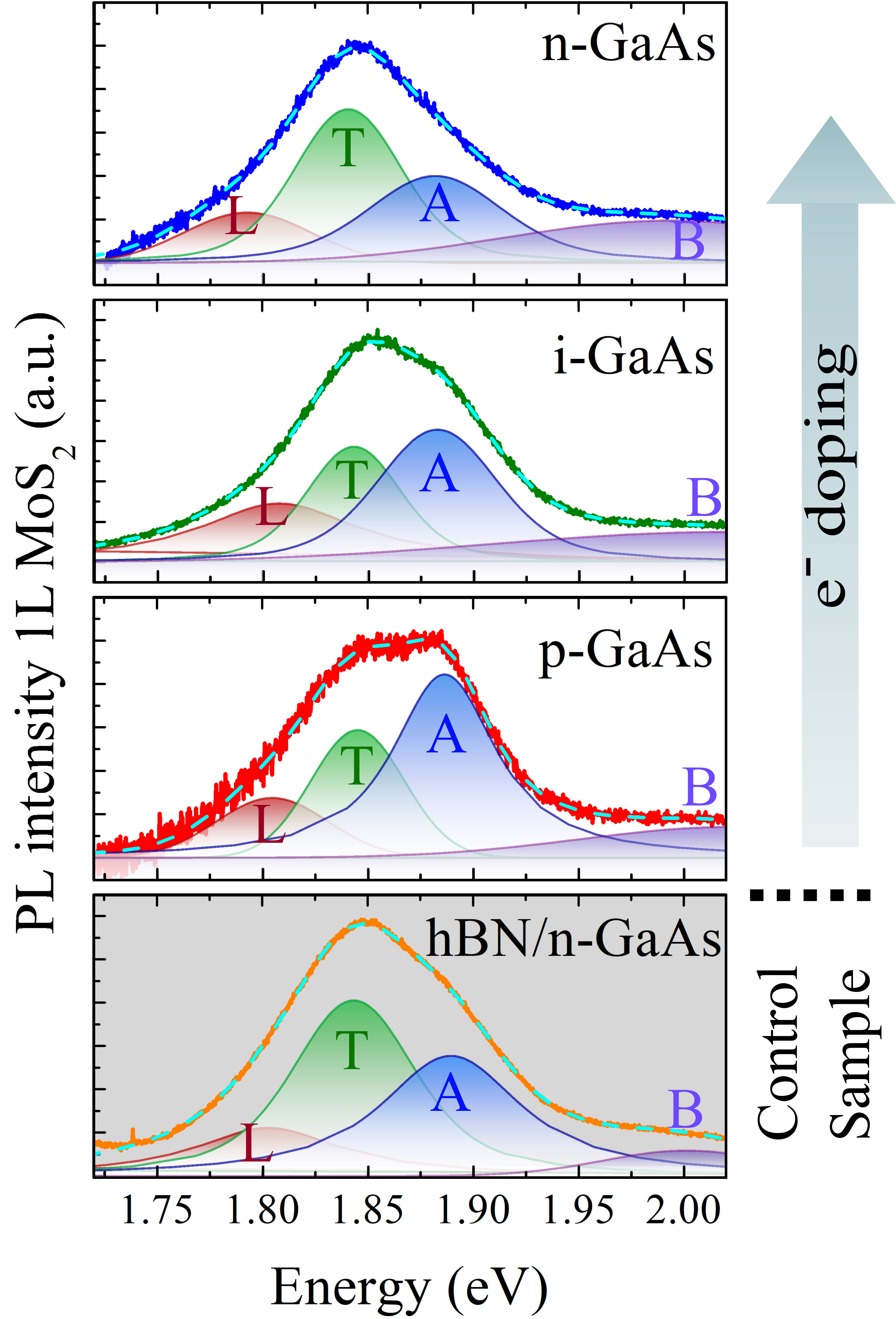}
    \caption{Peak decomposition of the photoluminescence spectra of ML-MoS$_2$ on different substrates. The peaks are contributions from: localized states (L), trions (T), A excitons (A), and B exciton (B) emissions. The solid lines are experimental data and the dashed lines represent the sum of the component peaks.}
    \label{Fig2:PL_fitts}
\end{figure}

To further understand the interaction between MoS$_2$ and GaAs in the heterostructures we decompose the PL spectra into peaks corresponding to the radiative recombination of different exciton species on ML-MoS$_2$. In Figure \ref{Fig2:PL_fitts} we present the PL spectra and their constituent peaks for all ML-MoS$_2$ on GaAs from the first set of samples and for the MoS$_2$/hBN/n-GaAs control sample. Four peaks with a Voigt lineshape were identified in the fitted spectra, the A and B exciton peaks, the trion peak (T), and a fourth peak (L), which has been previously assigned to the recombination of excitons bound to localized states \cite{Tongay2013, Saigal2016, Jadczak2017, Greben2020}.

The presence of a peak from recombination of trions, which are charged excitons, allows us to infer the existence of free charge, or an excess charge density, in ML-MoS$_2$. Exfoliated ML-MoS$_2$ are usually found to be intrinsically n-type \cite{Zhang2020, Singh2019}, having excess electrons in its conduction band. Thus, by comparing the integrated PL intensities of the trion peak, I$_T$, and of the A-exciton peak, I$_A$ (see Table \ref{Tab1:PL_T_A_ratio}) we can quantify the excess charge density comparatively among the samples and identify the relationship between the doping level of the substrate and the excess charge density on ML-MoS$_2$. A higher value of the ratio I$_T$/I$_A$ indicates higher excess charge density, as was observed for monolayers under electric gating \cite{Mak2013a, Greben2020, Pradeepa2020}.
Based on I$_T$/I$_A$ values (table \ref{Tab1:PL_T_A_ratio}) we can say that the excess charge density on ML-MoS$_2$ in our samples increases, depending on the substrate, in the following order: p-GaAs, i-GaAs, hBN/n-GaAs and n-GaAs. By assumption, the ML-MoS$_2$ in the control sample does not exchange charge with the substrate, therefore its I$_T$/I$_A$ is a measure of the isolated ML-MoS$_2$ excess electron density. The high contribution of trions in the control sample PL spectrum corroborates this assumption since it agrees with the already mentioned intrinsic n-type nature of exfoliated ML-MoS$_2$ samples, mostly related to sulfur vacancies \cite{Zhang2020,Singh2019}. Comparing the I$_T$/I$_A$ of the MoS$_2$/x-GaAs samples with the control sample we can infer that the n-GaAs substrate is the only one that transfers electrons to the monolayer, while inversely the i-GaAs and p-GaAs substrates receive electrons transferred from the MoS$_2$ monolayer.

The excess charge density on ML-MoS$_2$ is controlled by its Fermi level position. We expect that when the ML-MoS$_2$ and the substrate enter into contact they exchange charge carriers until their Fermi levels align, achieving an equilibrium state. This process may change the surface potential of GaAs causing some band bending but its Fermi level position is fixed by the bulk far from the surface. For ML-MoS$_2$, however, charge exchange will change its Fermi level position. Thus, the relations between I$_T$/I$_A$  among the samples give us a hint about the Fermi level change in ML-MoS$_2$ when it comes into contact with each substrate. Then, from a band alignment point of view, we may say that the Fermi level of ML-MoS$_2$, before contacting the substrate, is positioned somewhere between the Fermi level of the intrinsic and n-doped GaAs substrates. Nevertheless, from the I$_T$/I$_A$ connections alone, we cannot determine the band alignments for the different heterojunctions.

\begin{table}[h!]
\begin{center}
\begin{tabular}{lccc}
\hline
\hline
Sample	 & I$_A$ &  I$_T$ & I$_T$/I$_A$  \\
\hline
MoS$_2$/n-GaAs & 19.66 & 26.25 & 1.33 \\

MoS$_2$/i-GaAs & 27.86 & 19.12 & 0.69\\

MoS$_2$/p-GaAs & 37.13 & 16.31 & 0.44\\
%\rowcolor[gray]{0.9} 
MoS$_2$/hBN/n-GaAs & 666.82 & 745.26 & 1.12\\
\hline
\hline
\end{tabular}
\caption{Integrated photoluminescence intensities of the A exciton, I$_A$, and the trion, I$_T$, emission peaks of ML-MoS$_2$ in each sample, in arbitrary units, and their ratio, I$_T$/I$_A$.} \label{Tab1:PL_T_A_ratio}
\end{center}
\end{table}

In order to elucidate the band offsets of the three ML-MoS$_2$/x-GaAs heterojunctions, we used Scanning Kelvin Probe Microscopy (SKPM), which measures the contact potential difference (CPD) between the cantilever tip of an atomic force microscope and the surface of the sample \cite{Nonnenmacher1991, Melitz2011}. In the biased tip configuration, which we used for the SKPM measurements, by measuring the CPD and knowing the work function of the tip, $\phi_{tip}$, it is possible to determine the surface work function of the sample, $\phi_{samp}$, through the relation $e \cdot CPD=\phi_{tip}-\phi_{samp}$, where $e$ is the electron charge. We performed the experiments under standard ambient conditions, which can affect the precision of the specific values. Nevertheless, all uncertainties affect all samples equally, and we can confidently extract relationships between the surface work functions of the different materials in each sample measured.

\begin{figure}[b!]
    \centering
    \includegraphics[width=\linewidth]{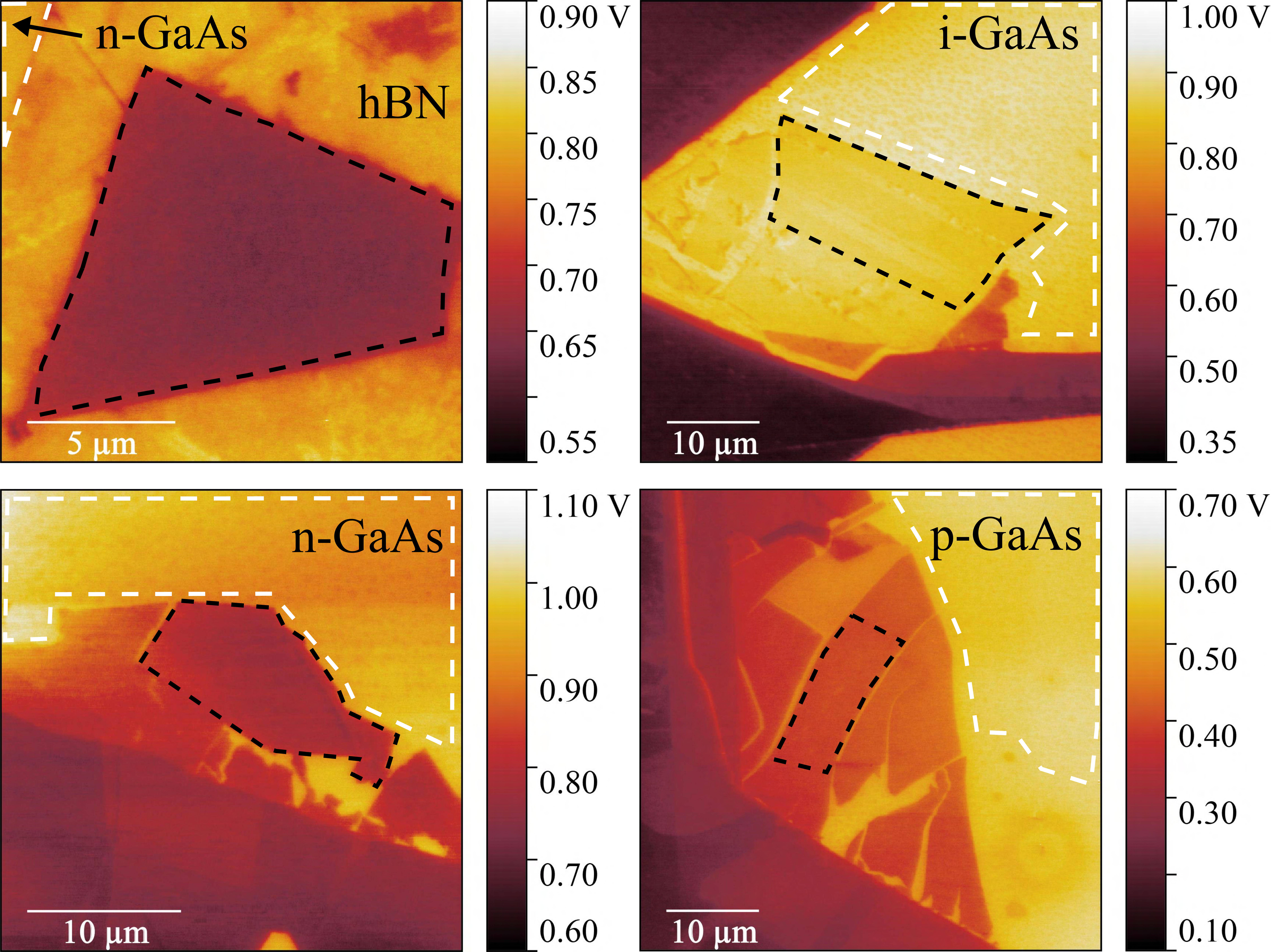}
    \caption{Contact potential difference maps obtained by SKPM of the studied heterojunctions. ML-MoS$_2$ (substrate) analyzed areas are delimited by black (white) dashed lines. The type of substrate is indicated in each map in a region of the image that corresponds to the substrate.}
    \label{Fig3:SKPM}
\end{figure}

\begin{figure*}[ht!]
	\includegraphics[width=\linewidth]{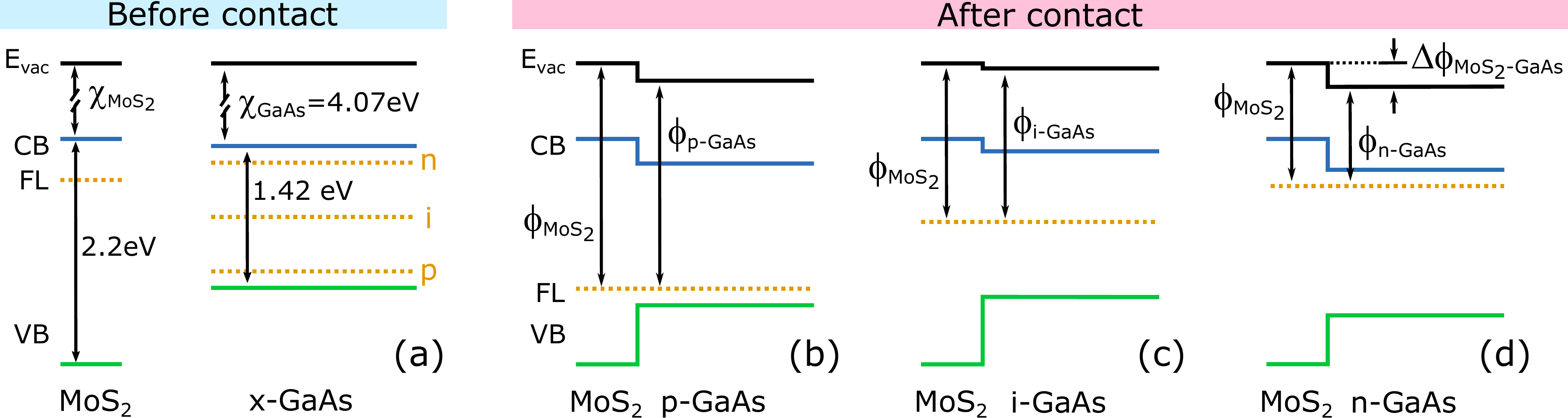}
	\caption{\label{Fig4:Bandoffset}Schematic band offsets of ML-MoS$_2$ and x-GaAs before contact (a), and after ML-MoS$_2$/x-GaAs heterojunction formation (b-d). E$_{vac}$, CB, FL, VB, $\chi$, and $\Phi$ represents the vacuum level, the bottom of the conduction band, the Fermi level, the top of the valence band, the electron affinity, and the work function respectively.}
\end{figure*}

To extract the CPD at each material we used the mean value of homogeneous areas of the monolayers, shown in Figure \ref{Fig3:SKPM} by dashed black lines, and the clean areas at each x-GaAs substrate, shown by dashed white lines in the figure. Optical images and sample details are shown in the SI. Therefore, it is possible to determine the difference between the work functions of the ML-MoS$_2$ and its corresponding substrate by the negative of the value of the CPD contrast, or  $\Delta\phi_{MoS_2-GaAs}= \phi_{MoS_2}-\phi_{GaAs}\ = e\ (V_{GaAs} - V_{MoS_2})$ (see Table \ref{Tab2:CPD_WF}). We observe that the obtained difference is positive for all samples, which indicates that the work function of MoS$_2$ is larger than the work function at the surface of GaAs in all samples.

To relate the work function of a material with its conduction and valence band edges we need to know the electron affinity $\chi$ and band gap E$_g$ of the material. The GaAs parameters are well established in literature: $\chi _{GaAs}=4.07~$eV and $E_{g,GaAs}=1.42~$eV \cite{Sze2006}. For ML-MoS$_2$, reports in the literature have a range of $\chi _{MoS_2}=$ 3.74 - 4.1 eV \cite{Keyshar2017, Guo2016} and the bandgap will suffer modulations owing to the dielectric screening from the environment, which in our samples should imply a value of $E_{gMoS_2} \sim 2.2~$eV considering the dielectric constant of GaAs as $\kappa _{GaAs}$= 12.88 \cite{Ryou2016}. To propose a band alignment for our heterojunctions we will consider $\chi _{MoS_2}=4.0~$eV, which was the value used in other works on MoS$_2$/GaAs \cite{Lin2015, Xu2016, Padma2019} and the electronic bandgap.

\begin{table}[h!]
\begin{center}
\begin{tabular}{lcc}
\hline
\hline
Sample   & $\Delta\phi_{MoS_2-GaAs}$ (eV) \\
\hline
%\rowcolor[gray]{0.8}
MoS$_2$/n-GaAs  &   0.23 $\pm$ 0.04 \\

MoS$_2$/i-GaAs  &   0.05 $\pm$ 0.05 \\
%\rowcolor[gray]{0.8}

MoS$_2$/p-GaAs  &   0.22 $\pm$ 0.03 \\

MoS$_2$/hBN/n-GaAs  &   0.14 $\pm$ 0.01 \\
\hline
\hline
\end{tabular}
\caption{Work function difference between ML-MoS$_2$ and GaAs extracted from the SKPM maps shown on Figure \ref{Fig3:SKPM}.} \label{Tab2:CPD_WF}
\end{center}
\end{table}

Since the position of the conduction band can be described as $E_c=\phi - \chi$, with respect to the Fermi level, we approximate the difference in the conduction band edge between the MoS$_2$ layer and the x-GaAs substrate by $\Delta E_c=E_{c,MoS_2}-E_{c,GaAs}=\Delta\phi_{MoS_2-GaAs}+\Delta\chi_{GaAs-MoS_2}$. As both quantities are positive, the conduction band edge of MoS$_2$ is always at a higher energy than that of GaAs, with their Fermi levels aligned.

Figure \ref{Fig4:Bandoffset} presents schematically the band offsets we propose for the ML-MoS$_2$/x-GaAs heterojunctions based on our analysis of the PL and SKPM results. In Figure \ref{Fig4:Bandoffset}a we present the band edges and Fermi levels of each material before contact. Fermi levels are represented by the yellow dotted lines and, in GaAs, are labeled n, i, and p for the type of substrate doping. As inferred from the PL I$_T$/I$_A$ analysis, we position the Fermi level of MoS$_2$ between those of i-GaAs and n-GaAs. Heterojunction band alignments after contact are shown in Figures \ref{Fig4:Bandoffset}b, \ref{Fig4:Bandoffset}c and \ref{Fig4:Bandoffset}d. According to our proposal, ML-MoS$_2$ and GaAs form type I heterojunctions for all GaAs doping levels studied.

The SKPM data does not give a quantitative, exact value of the conduction band offset in the heterojunction (see SI for more details on the technique). Nevertheless, there is a clear indication that the steps in conduction band at the junction are of comparable magnitude for all three types of GaAs substrates. After establishing the conduction band step at the junction, the position of the Fermi level is set by the doping of the GaAs substrate, according to the assumption that the Fermi level is pinned down by the bulk of the material. This determines the position of the Fermi level in the MoS$_2$ side of the junction.

As the estimated differences in work function obtained from SKPM are between the surface work functions, the band alignments we present in Figure \ref{Fig4:Bandoffset} assume that the surface work function of GaAs is the same as its bulk work function, or that the GaAs bands are flat. We prefer not to speculate on the curvature of the bands inside GaAs because our experiments do not provide sufficient evidence to support it. This means that, although at the interface the band positions we proposed should be correct, the curvature of the GaAs bands may change as one moves from the surface to the bulk, which means that the Fermi level positioning should also be reexamined. Therefore, we propose the band alignments in Figure \ref{Fig4:Bandoffset} as a first approximation, to contribute to the discussion and analysis of the surface and charge dynamics in these 2D/3D heterostructures and we expect to instigate other works aiming to elucidate the shape of the bands inside GaAs on these types of junctions, since band bending can affect the operation of devices based on them.

Most of the work done on MoS$_2$/GaAs junctions so far employ n-doped GaAs \cite{Xu2016, Lin2015, Padma2019, Sarkar2020, Jia2019}. Nearly all of these works propose a type II band alignment for MoS$_2$/n-GaAs. That is not in complete disagreement with our proposal, since the transition to type II alignments for the MoS$_2$/n-GaAs junctions would only imply that the conduction band step is larger than the one we estimated, which is based in comparisons of the experimental data for the three types of MoS$_2$/x-GaAs junction and the control sample. Furthermore, it is worth pointing out that the devices studied in these other works were built with MoS$_2$ produced by chemical vapour deposition (CVD) \cite{Xu2016, Lin2015, Padma2019}, thermal decomposition \cite{Jia2019} or solution processing \cite{Sarkar2020, Zhang2017}, while we used exfoliated ML-MoS$_2$. That could be relevant since it is well known that the defects, and thus doping, of MoS$_2$ monolayers obtained through each method can be quite different.

Our proposed type I band alignments for all the studied heterojunctions implies that the mechanism behind the quenching of the ML-MoS$_2$ photoluminescence in the heterojunctions should be the transfer of excitons from MoS$_2$ to GaAs and not exciton dissociation through the junction. Additionally, the different Fermi level positions in ML-MoS$_2$ on different substrates allows us to explain the variations in the relative intensity of the emission from the trion and the A exciton that were observed. 

%\section*{Conclusions}

In conclusion, we presented the photoluminescence spectra of monolayers of MoS$_2$ on commercial GaAs substrates with different doping levels. The results revealed an important reduction of the PL intensity of the monolayers, when compared with a control sample. In addition, the spectra presented a dependence of the ratio of the trion to exciton emission intensities on the doping level of the substrate. This behavior evidences different ammounts of excess charge in the single layers related to a charge exchange process with their substrates.
Scanning Kelvin probe microscopy measurements provided an estimation of the difference in work function between the materials in the heterojunctions and allowed us to propose a type I band alignment for all MoS$_2$/x-GaAs heterojunctions we studied. Our proposal is consistent with the analysis of the photoluminescence measurements and suggests exciton migration as the main mechanism behind the PL intensity reduction. The results reported here contribute to the understanding of the charge transfer processes in 2D/3D semiconductor heterojunctions which are of central importance for the implementation of the next generation of electronic and optoelectronic devices.

\section*{Supplementary Material}
See Supplementary Information for additional experimental details regarding the fabrication, characterization and spectra processes of the heterojunctions.

\begin{acknowledgments}
An important part of the work reported here was done at the LCPNano laboratory at UFMG. We thank Freddie Hendriks for the calculations of thin-film interference effects shown in the SI. This work was financially supported by the Brazilian funding agencies CNPq, FAPEMIG and the Coordenação de Aperfeiçoamento de Pessoal de Nível Superior - Brasil (CAPES).% -  Finance Code 001.
\end{acknowledgments}

\section*{Data Availability Statement}

The data that support the findings of this study are available from the corresponding author upon reasonable request.

\section*{Conflict of interest}

The authors have no conflicts to disclose.

%\section{Appendixes}

\section*{References}
\bibliography{aipsamp}% Produces the bibliography via BibTeX.

%--------------------SUPPLEMENTARY -----------------------------------
\onecolumngrid
\clearpage

\section*{\hfil \hfil \large Supplementary Information}
\renewcommand{\figurename}{Figure S}
\section{Sample fabrication and characterization}

\begin{figure}[ht!]
    \centering
    \includegraphics[width=\linewidth]{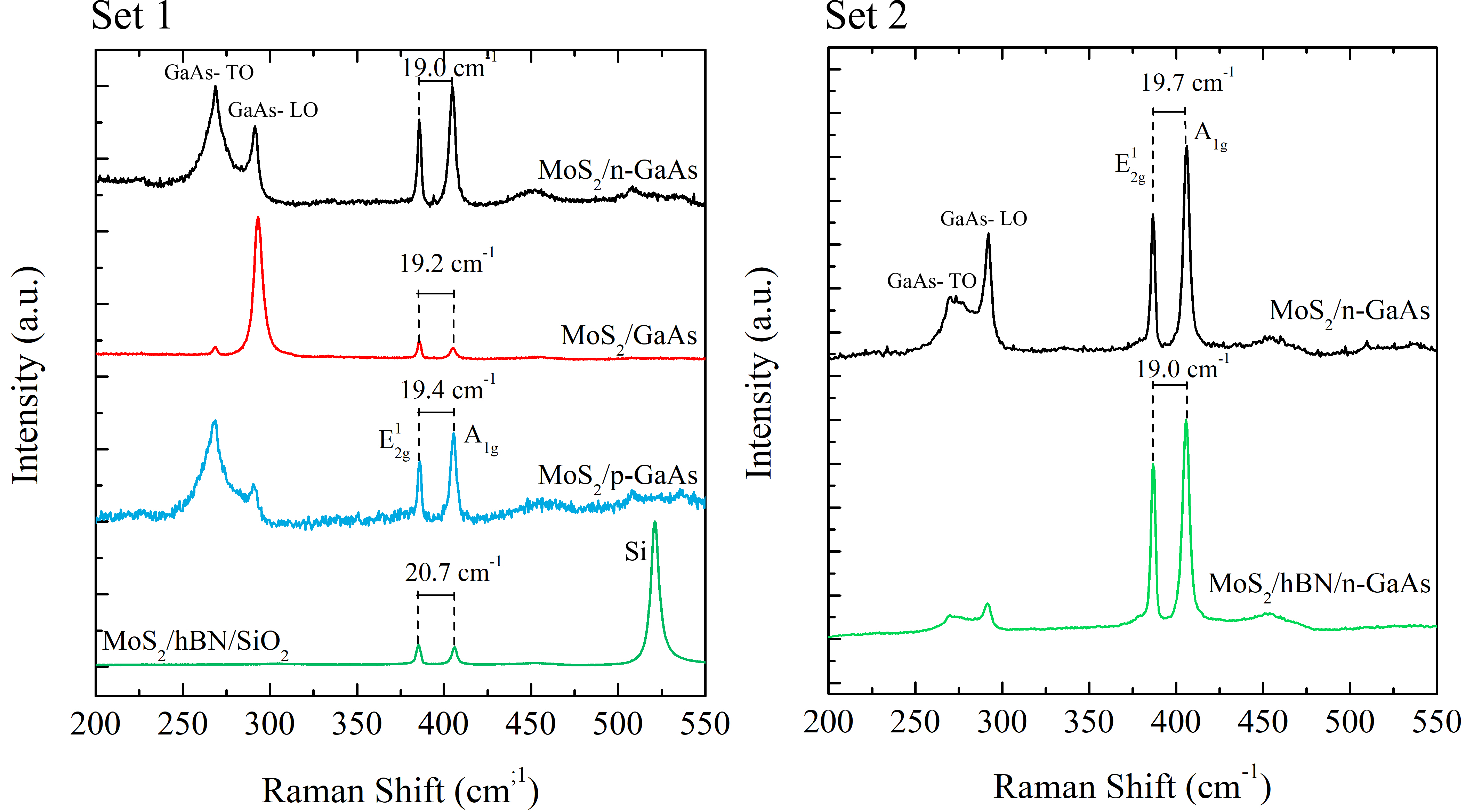}
    \caption{Raman spectra of ML MoS$_2$ on the studied substrates. All  samples show the Raman shift separation between the MoS$_2$ A$_{1g}$ and E$_{2g}$ peaks that is characteristic of MoS$_2$ single layers. Optical phonon modes from the substrates can also be seen in the spectra.}
    \label{FigSI1:Raman}
\end{figure}

%\raggedright \setlength\parindent{30pt}
The samples were studied in two sets. The first set was composed of a control sample of monolayer MoS$_2$ on hBN/SiO$_2$/Si substrate (MoS$_2$/hBN/SiO$_2$) and three samples of ML-MoS$_2$ on GaAs with different doping:  MoS$_2$/i-GaAs, MoS$_2$/p-GaAs and MoS$_2$/n-GaAs. The second set is composed of two samples, one ML-MoS$_2$ on n-GaAs and one ML-MoS$_2$ control sample on hBN/n-GaAs substrate (MoS$_2$/hBN/n-GaAs). The second set of samples allowed us to confirm the insulating quality of hBN on the control samples as well as to verify the reproducibility of the results obtained for ML-MoS$_2$ on n-GaAs. 

To fabricate the samples, we cleaned the Si/SiO$_2$ substrate via isopropyl alcohol and acetone baths for 5 minutes each, followed by high-pressure N$_2$ gas to eliminate any possible impurity. In addition, GaAs substrates were also cleaned by Ar plasma for 15 minutes at 0.300 Torr and 250 W RF power after the wet cleaning. Monolayers of MoS$_2$ (ML-MoS$_2$) were mechanically exfoliated and transferred to the substrates through the all-dry viscoelastic stamp method \cite{Castellanos-Gomez2014}. Similar processes were used to exfoliate and transfer the hBN bulk to the Si/SiO$_2$ and n-GaAs substrates. 

To confirm the monolayer nature of the MoS$_2$ in our samples , we used Raman spectroscopy, with a 532 nm laser at 0.5 mW. In monolayer MoS$_2$, the separation in frequency of the well-known $A_{1g}$ and $E^{1}_{2g}$ Raman modes, should be close to, or smaller than, 19 cm$^{-1}$ \cite{Lee-Heinz-Hone-ACSNano2010,Li-Zhang-Advanced-Funct-Materials2012}. The obtained spectra are presented in Figure S \ref{FigSI1:Raman}. In addition to the mentioned peaks, we can see the TO and LO Raman modes of the GaAs substrates as well as the Si peak of the control sample in set 1.

\begin{figure}[h!]
    \centering
    \includegraphics[scale=0.19]{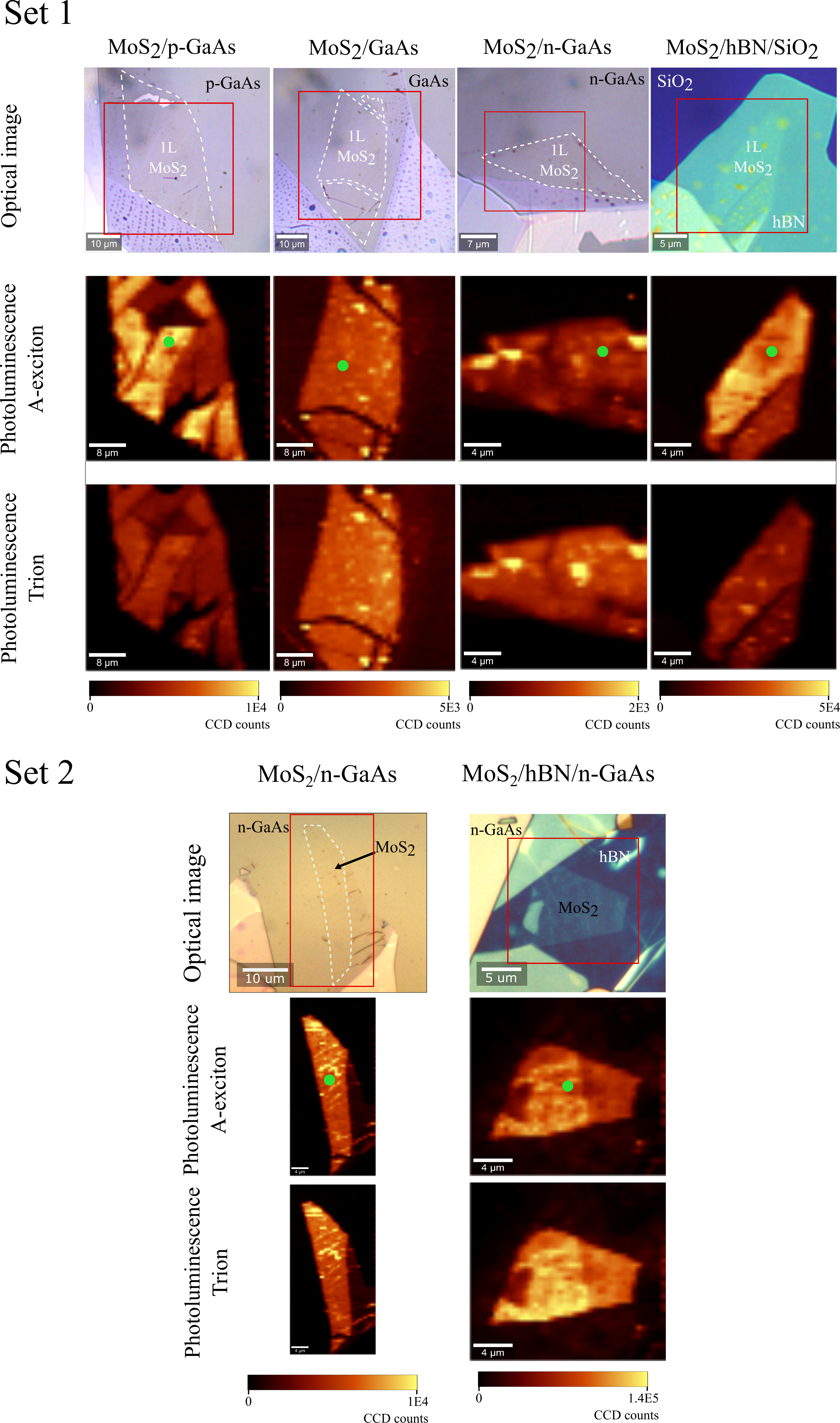}
    \caption{Optical images and photoluminescence maps of the studied samples. }
    \label{FigSI2:Samples}
\end{figure}

Photoluminescence spectra were measured in a WITec \textit{alpha 300A} experimental setup with a laser power of 300 $\mu$W, at a wavelength of 457 nm. The spectra presented in Figures 1 and 2 of the main document were acquired as the average of 4 consecutive measurements accumulated for 40 seconds, at the points shown by the green dots in Figure S\ref{FigSI2:Samples}. The latter presents the optical image of the samples and two sets of PL maps, obtained by integrating the PL spectrum at each point in a region of 20 meV width, with the center in 1.88 eV for the middle panel, and 1.84 eV for the bottom panel. The color scale is the same for both panels of each sample and is shown at the bottom of the figure. The PL maps in Figure S\ref{FigSI2:Samples} were acquired with accumulation time of approximately one second per spectra and obtained with spatial steps of 0.5 $\mu$m. Raman spectra were measured with acquisitions of 1 minute. PL spectra and Raman spectra were obtained at different representative points of the sample and the PL maps were done afterwards, to minimize photodoping effects.

\clearpage
\section{SKPM measurements}

There are several types of SKPM methods (amplitude-modulation (AM), frequency-modulation (FM), homodyne-detection, heterodyne-detection and others) \cite{1Ulrich2005,2Zong2013,3Garrett2016,4Axt2018}. Some good reviews on this subject are presented in references \cite{1Ulrich2005} to \cite{4Axt2018} and references therein. As a general rule, AM-based SKPM in ambient conditions yields qualitative surface potential values, whereas FM-based SKPM is employed when quantitative surface potential values are needed. This is mainly due to the strong influence of the cantilever (and not only the tip apex) on the SKPM signal in AM-based methods \cite{1Ulrich2005,2Zong2013,3Garrett2016,4Axt2018}. In the present work, the SPM system used is capable of conventional AM-based SKPM only and, thus, the yielded results should be considered qualitative.

To work within the limitations of the method and still be able to make reliable comparisons between samples we elect the control sample from the second set, with a substrate of n-GaAs below the hBN flake, as the better control sample for SKPM measurements. It allows us to have a similar cantilever-substrate interaction effect on the SKPM measurements.

In particular, our measurements were conducted on Bruker Multimode 8 with a Nanoscope V controller, at normal atmospheric pressure and room temperature. 
%\newpage
\subsection{Atomic force microscopy}
\begin{figure}[ht!]
    \centering
    \includegraphics[scale=0.27]{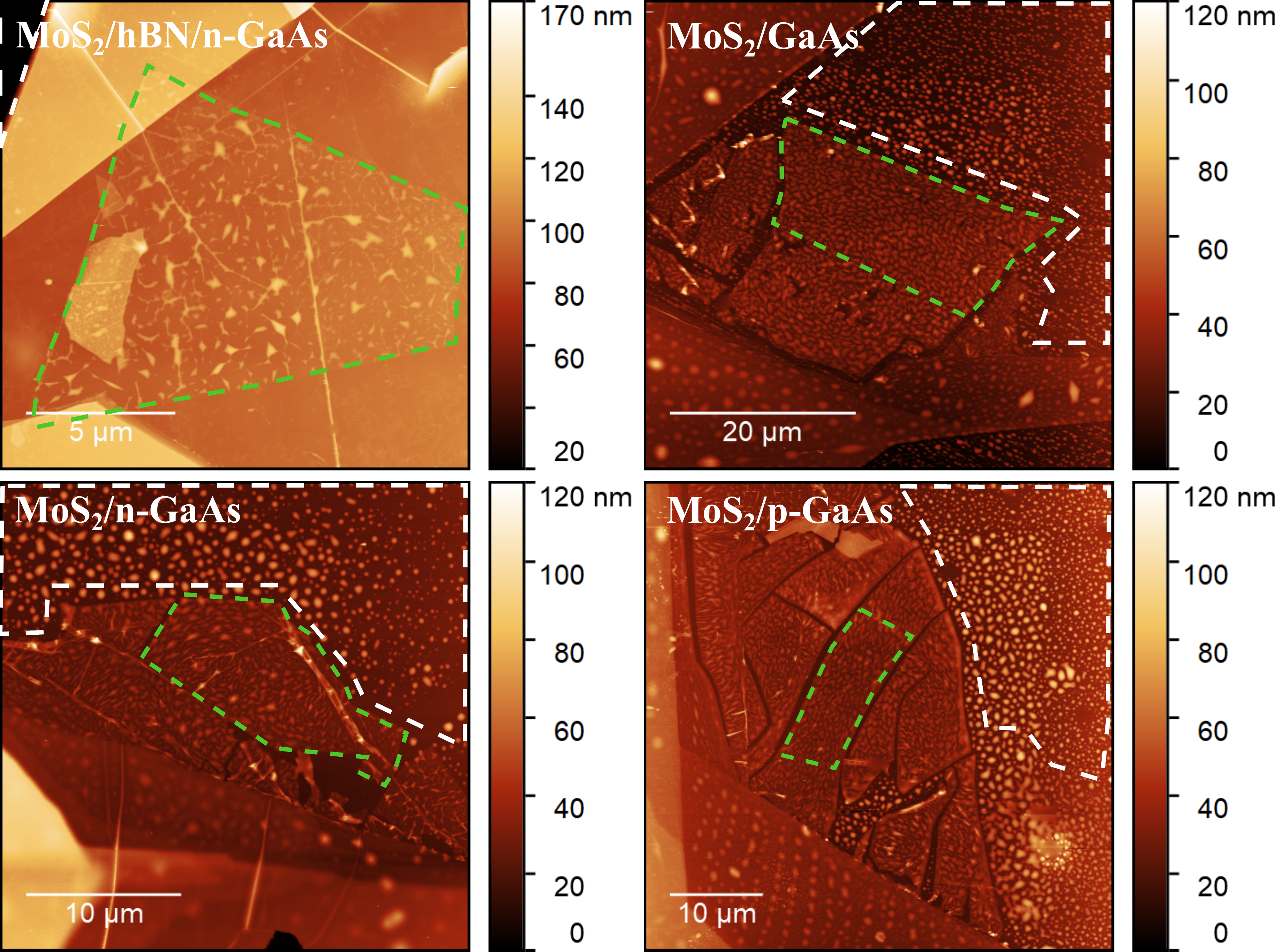}
    \caption{Atomic force microscopy topographic images of the studied samples. ML-MoS$_2$ (substrate) analyzed areas are delimited by green (white) dashed lines.}
    \label{FigSI3:AFM}
\end{figure}

%\clearpage
\section{Photoluminescence}

\begin{figure}[h!]
    \centering
    \includegraphics[scale=0.45]{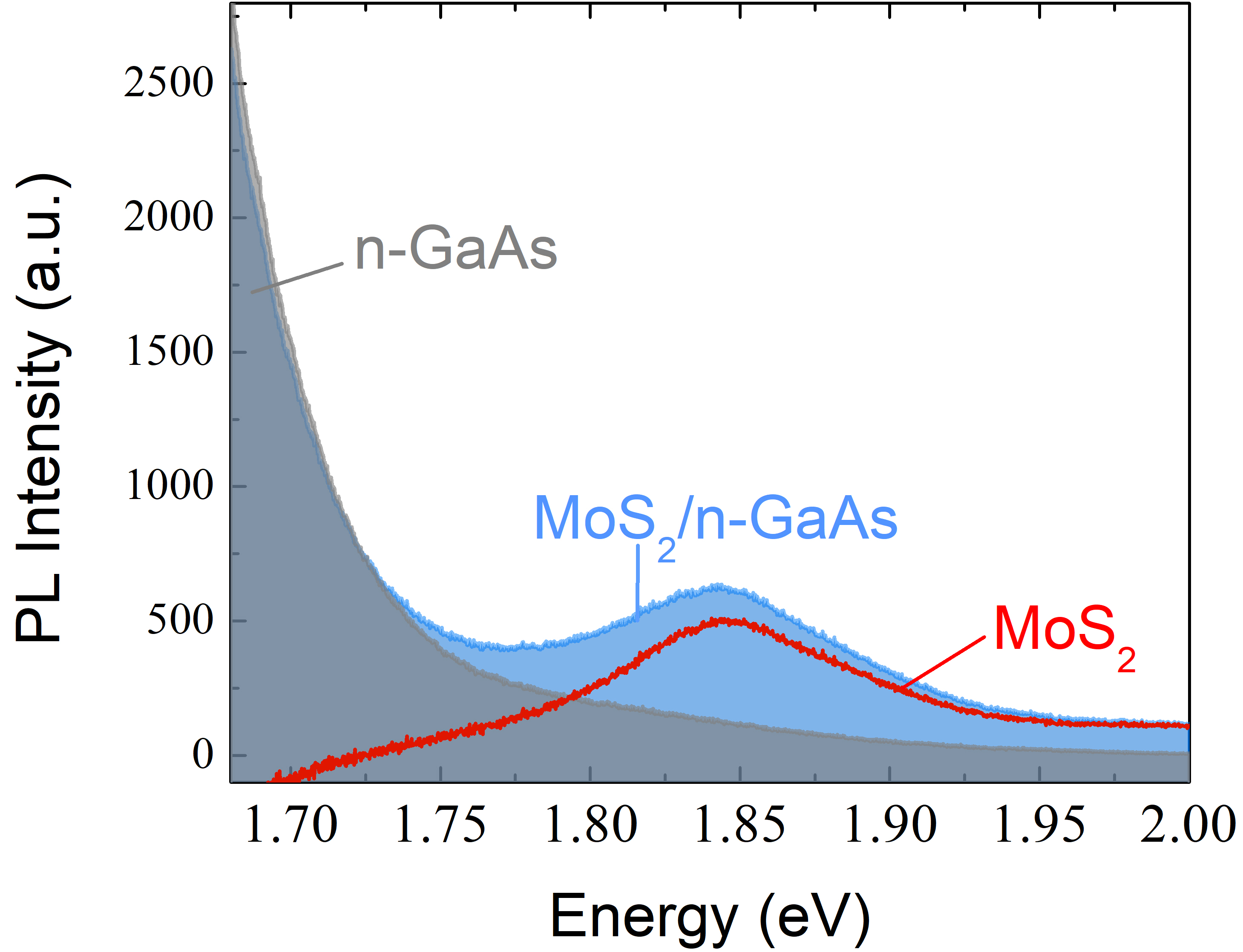}
    \caption{Photoluminescence (PL) spectra substraction of MoS$_2$/n-GaAs. In gray the PL from the bare substrate, in blue the PL measured on the monolayer MoS$_2$ on the n-GaAs substrate, in red the resulting PL of gray minus blue. }
    \label{FigSI4:PL_Subs}
\end{figure}

Since GaAs and MoS$_2$ have direct bandgaps with relatively close energy values and owing to the atomic thickness of the MoS$_2$, the photoluminescence spectra obtained
from MoS$_2$/GaAs heterojunctions are composed of emissions from both materials. By subtracting the emission of the substrate from the heterojunction spectra we are left with the ML-MoS$_2$ emission. For the MoS2/hBN/SiO2 control sample, the substrate does not have a strong PL signal and the PL spectrum measured on the heterostructure is already the ML-MoS$_2$ emission. Figure S \ref{FigSI4:PL_Subs} illustrates the separation process for the spectrum of MoS2 on n-GaAs.  

\newpage
\subsection{Photodoping effect}

\begin{figure}[h!]
    \centering
    \includegraphics[width=\linewidth]{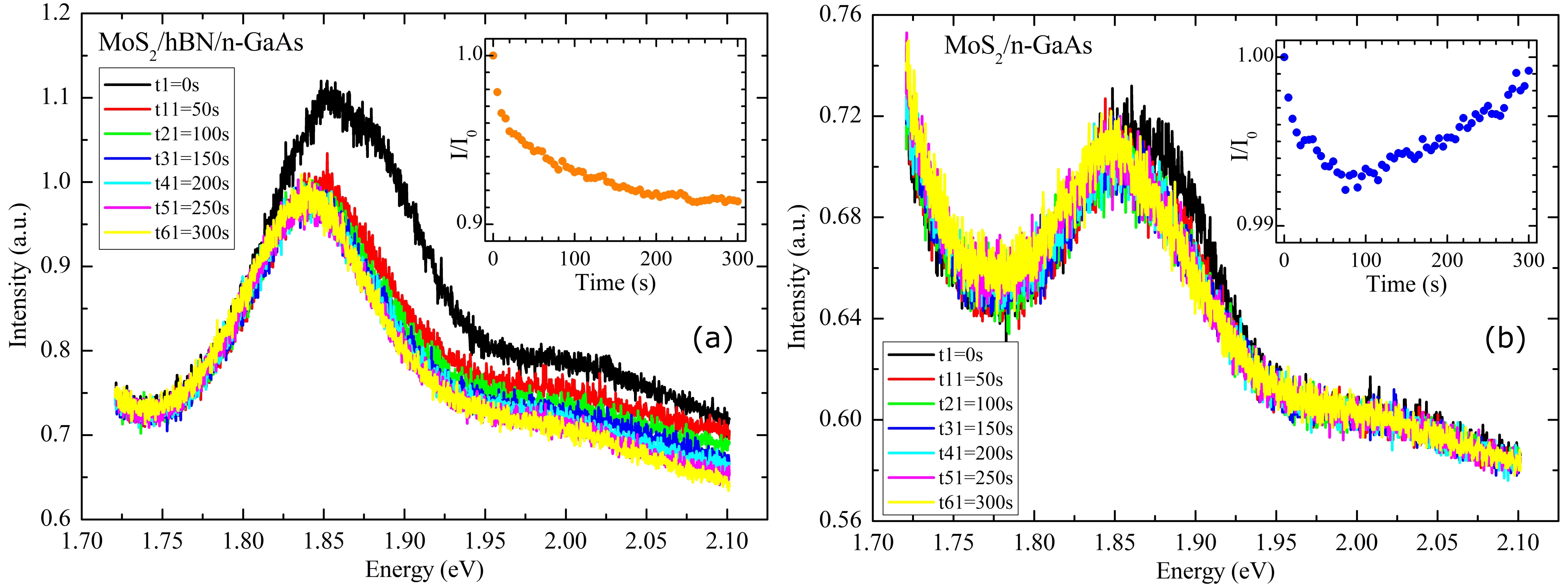}
    \caption{Photodoping effect investigation. Spectra obtained by continuously illuminating the sample with the 457 nm laser, with power of 300 $\mu$W and accumulation time of 1 s for the MoS$_2$/hBN/n-GaAs in (a) and 3 s for MoS2/n-GaAs in (b). Spectra were acquired every 5 s, but we only show the spectra obtained at times 0, 50, 100, 150, 200, 250 and 300 s, for clarity. Insets show the time (exposure) dependence of the relative integrated intensity, I/I0, which is the integrated intensity of each spectrum divided by the integrated intensity of the first spectrum at t = 0 s. The intensity variation along the measurement is less than 10$\%$ for MoS$_2$/hBN/n-GaAs, and less than 1$\%$ for MoS$_2$/n-GaAs. The plots show raw data where MoS$_2$ and GaAs emissions were not separated, the tail of the GaAs emission peak (centered near 1.4 eV) is clear in the low energy signal of the MoS$_2$/n-GaAs sample data.}
    \label{FigSI5:PL_photodoping}
\end{figure}
%\newpage
\subsection{Interference effects}

\begin{figure}[h!]
    \centering
    \includegraphics[scale=0.7]{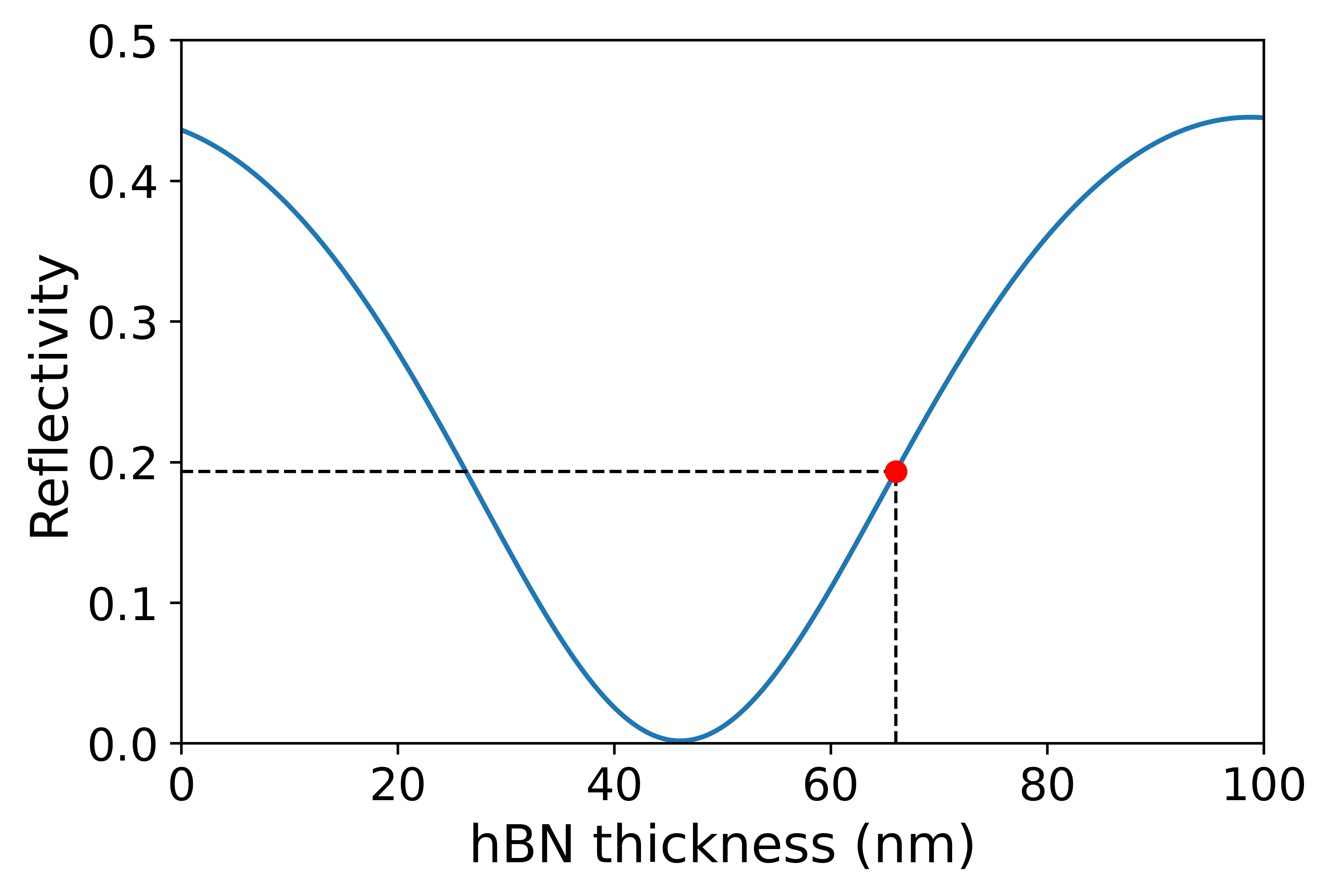}
    \caption{Interference effect on the reflectivity of a MoS$_2$/hBN/GaAs stack as a function of the hBN thickness for the laser excitation wavelength of $\lambda$ = 457 nm.}
    \label{FigSI6:Reflectivity}
\end{figure}

To evaluate the magnitude of cavity effects on the reflectivity of our samples, we performed transfer matrix calculations of the reflectivity for our samples and find that no large changes should be expected. Figure S\ref{FigSI6:Reflectivity} shows the reflectivity of a MoS$_2$/hBN/GaAs stack as a function of the hBN thickness, for the laser excitation wavelength. At the thickness determined by atomic force microscopy of the hBN layer on our samples, t$_{hBN}$ = (66 $\pm$ 5) nm, the reflectivity of the whole stack is just below 0.2, as compared to 0.43 for a sample without hBN. Thus, the excitation laser absorption does not change by more than 20$\%$. A similar estimate for the interference effects at the luminescence wavelength indicates a destructive interference with the addition of the hBN layer. Therefore, cavity effects do not explain the large change in PL intensity between the control samples and MoS$_2$/GaAs.

\end{document}